\documentclass[sigconf]{acmart}

\usepackage[dvipsnames]{xcolor}
\usepackage{hyperref}       
\usepackage{url}
\usepackage{tcolorbox}
\usepackage{multirow}
\usepackage{subcaption}
\AtBeginDocument{%
  }

\setcopyright{acmlicensed}
\copyrightyear{2018}
\acmYear{2018}
\acmDOI{XXXXXXX.XXXXXXX}
\acmConference[Conference acronym 'XX]{Make sure to enter the correct
  conference title from your rights confirmation email}{June 03--05,
  2018}{Woodstock, NY}
\acmISBN{978-1-4503-XXXX-X/2018/06}

\begin{document}

\title{HGCRS: Hypergraph-Enhanced Generative Conversational Recommender System}

\author{Sixiao Zhang}
\affiliation{%
  \institution{Nanyang Technological University}
  \country{Singapore}
}
\email{sixiao001@e.ntu.edu.sg}

\author{Mingrui Liu}
\affiliation{%
  \institution{Nanyang Technological University}
  \country{Singapore}}
\email{mingrui001@e.ntu.edu.sg}

\author{Cheng Long}
\authornote{Corresponding author.}
\affiliation{%
  \institution{Nanyang Technological University}
  \country{Singapore}}
\email{c.long@ntu.edu.sg}

\renewcommand{\shortauthors}{Trovato et al.}

\begin{abstract}
Recent conversational recommender systems have begun to unify recommendation and response generation within autoregressive models, with semantic IDs providing a natural representation for generative recommendation. However, relational knowledge is typically organized around item-level representations, which do not naturally align with the compositional SIDs. This mismatch makes it difficult for unified CRSs to effectively exploit structured catalog relations. We propose HGCRS, a hypergraph-enhanced generative conversational recommender system that bridges this gap with an SID--entity hypergraph. Each item forms a hyperedge connecting its SID codes and attribute entities, and a type-aware hypergraph encoder incorporates relational information into generative item selection. HGCRS further introduces a mode--item--response protocol for jointly modeling recommendation-bearing and response-only turns. Experiments on ReDial and INSPIRED demonstrate strong recommendation performance while maintaining competitive response generation quality.
\end{abstract}




\maketitle

\section{Introduction}
\label{sec:introduction}

Conversational recommender systems (CRSs) support item discovery through multi-turn natural-language interactions \cite{li2018towards}. Users can express preferences, ask about items, respond to suggestions, and refine their requests over the course of a dialogue. The conversation therefore provides an evolving context for identifying relevant items and generating responses that address the user's current needs. Recommendation and response generation are inherently coupled: item selection depends on preferences expressed through language, while the response communicates recommendations, explains them, and elicits further feedback. Effectively integrating these two capabilities has consequently become a central direction in CRS research \cite{zhou2020improving,wang2022towards}.

Earlier CRSs generally adopt a modular architecture in which recommendation and dialogue generation are handled by separate components \cite{chen2019towards,zhou2020improving,wang2022towards}. A recommendation module infers user preferences and ranks candidate items, while a dialogue module generates responses based on the conversational context and the retrieved recommendations. Although effective, such pipelines only loosely couple the two tasks: recommendation and response generation are often optimized with different objectives, and information between them is exchanged through intermediate representations or retrieved items. This separation limits the extent to which language understanding, item selection, and response generation can mutually inform one another. Recent work therefore moves toward \emph{unified generative recommender systems}, in which item recommendation is incorporated directly into a language model and recommendation targets are generated together with natural-language responses \cite{doh2025talkplay,de2026unified}. Semantic IDs (SIDs) \cite{NEURIPS2023_20dcab0f} provide a particularly suitable item representation for this formulation. Instead of assigning an independent token to every item, an SID represents an item as a short sequence of reusable code tokens. Item selection can consequently be formulated under the same autoregressive next-token objective as response generation, allowing conversational understanding, recommendation, and language generation to be more tightly integrated within a single model.

Beyond conversational context itself, however, catalog relations provide an important source of information for recommendation. A dialogue may reveal that a user likes a particular director, actor, creator, or previously discussed item, but the dialogue text alone does not explicitly specify which other items are connected to that entity or how candidate items are related through shared attributes. Knowledge graphs (KGs) provide precisely this complementary information by explicitly representing relations among items and their associated entities. Their usefulness has been demonstrated extensively in modular CRSs \cite{chen2019towards,zhou2020improving,wang2022towards}, where KG representations connect entities mentioned in dialogue to candidate items and provide relational signals for recommendation. While unified generative CRSs provide tighter integration between recommendation and language generation, they lack an equally natural mechanism for making such structured entity--item relations directly available to the compositional SID representations.

The difficulty is not merely to attach an existing KG encoder to the generative model, because conventional knowledge-aware recommendation and SID-based generation operate at different representation granularities. In conventional knowledge-aware CRSs, each item is represented as an atomic candidate with a corresponding graph node, and prediction directly scores that same item. The object represented by the graph and the target selected by the recommender therefore share the same granularity. This correspondence disappears in SID-based generative recommendation. An item is no longer selected through one atomic item representation, but by generating a sequence of SID code tokens. Although the complete SID uniquely identifies an item, its individual codes are intentionally shared across multiple items. An item-level graph embedding therefore has no one-to-one counterpart among the representations actually used by the generator. Simply retaining conventional item nodes does not determine how relational evidence should influence these shared code representations. The key challenge is therefore to make catalog relations directly available to SID-based generative recommendation.

To address this challenge, we propose \emph{Hypergraph-Enhanced Generative
Conversational Recommender System} (\emph{HGCRS}), which grounds item
generation in an SID--entity hypergraph. Rather than representing each item
as an atomic graph node, HGCRS represents it as a hyperedge connecting its
SID codes with its associated entities. A hypergraph provides a natural formulation for this structure, as each item
jointly associates multiple position-specific SID codes with multiple typed
entities, while these SID codes and entities can be shared across items. The hypergraph therefore aligns relational
modeling directly with the SID representations: SID nodes correspond to the shared code tokens, while entity nodes represent catalog attributes that can be
mentioned in dialogue. Shared SID-code and entity nodes link multiple item
hyperedges, allowing relational information to propagate across items. A
type-aware hypergraph encoder further distinguishes different SID positions
and entity roles during message passing. Its representations are injected
through gated residual connections into the LLM's entity and SID input
embeddings as well as its SID output embeddings, enabling relational
information to influence both dialogue interpretation and generative item
selection.

HGCRS additionally adopts a structured mode--item--response generation protocol to accommodate the different assistant behaviors that arise in free-form multi-turn conversations. Each assistant turn is represented by a mode, an optional recommendation SID, and a natural-language response. The mode distinguishes turns that introduce a new recommendation from those that continue the conversation without a new item target, while constrained decoding ensures that generated SIDs correspond to valid items. This protocol explicitly separates the decision of \emph{whether to recommend} from \emph{what to recommend}, while representing both recommendation decisions and natural-language responses within a shared autoregressive sequence. As a result, recommendation-bearing and response-only turns can be modeled through a common generation interface while preserving their distinct conversational roles. Our contributions are summarized as follows:
\begin{itemize}
    \item We propose HGCRS, a hypergraph-enhanced generative conversational recommender system that enables SID-based generative recommendation to exploit structured catalog relations. HGCRS bridges the granularity mismatch between item-level relational representations and shared SID-code tokens through an SID--entity hypergraph and type-aware relational propagation.

    \item We introduce a structured mode--item--response generation protocol that explicitly distinguishes recommendation-bearing and response-only turns while jointly generating recommendation targets and natural-language responses within a single autoregressive model.

    \item Extensive experiments on ReDial and INSPIRED demonstrate that HGCRS consistently improves recommendation performance while maintaining competitive response generation quality. Further ablation studies validate the effectiveness of relational grounding, type-aware relational encoding, SID configuration, and structured generation.

\end{itemize}

\section{Related Work}
\label{sec:related_work}

\textbf{Conversational recommender systems.}
Most conventional CRSs adopt a modular architecture in which recommendation
and response generation are handled by separate components. Early methods exploit knowledge graphs to connect dialogue
entities with candidate items
\cite{chen2019towards,zhou2020improving,lin2023cola}, while later work
incorporates reviews, contrastive learning, and pretrained language models
\cite{lu2021revcore,zhou2022c2,wang2022recindial,wang2022towards}.
Recent systems further leverage LLMs for preference understanding,
reranking, or response generation
\cite{friedman2023leveraging,xi2024memocrs,zhu2025collaborative,li2026improving},
while retaining an explicit item-level recommendation stage. Such
architectures naturally accommodate structured entity--item relations because
both relational modeling and recommendation operate over atomic items.

\textbf{Unified generative recommendation.}
Recent work instead generates recommendation targets and natural-language
responses within a shared autoregressive model. LUMEN
\cite{su2025regen} jointly supports recommendation-related tasks and
narrative generation, NEO \cite{de2026unified} generates semantic-ID tokens
together with natural language under instruction-specified formats, and
TalkPlay \cite{doh2025talkplay} jointly models dialogue and item generation
for conversational music recommendation. These studies demonstrate the
feasibility of unified generative recommendation, but do not explicitly address how
structured catalog relations should interact with compositional item
representations. In particular, conventional relational models represent an
item atomically, whereas SID-based generators select it through multiple
shared code tokens.

\textbf{Generative recommendation with semantic IDs.}
Generative recommendation replaces candidate scoring with autoregressive item
identifier generation. TIGER \cite{NEURIPS2023_20dcab0f} represents items
with semantic IDs obtained through residual quantization, and subsequent work
improves discrete item representations through alternative quantization,
collaborative signals, collision handling, and multimodal information
\cite{deng2025onerecunifyingretrieverank,10.1145/3773771,10597986,
10.1145/3627673.3679692,zhai2025multimodalquantitativelanguagegenerative,
zhang2025multiaspectcrossmodalquantizationgenerative,
li2025bbqrecbehaviorbindquantizationmultimodal}.

HGCRS connects these directions by aligning structured catalog relations with
the shared SID-code representations directly used for generative item
selection, while jointly modeling recommendation and natural-language
response generation.

\section{Methodology}
\label{sec:methodology}

\begin{figure*}[t]
\centering
\includegraphics[width=1.0\textwidth]{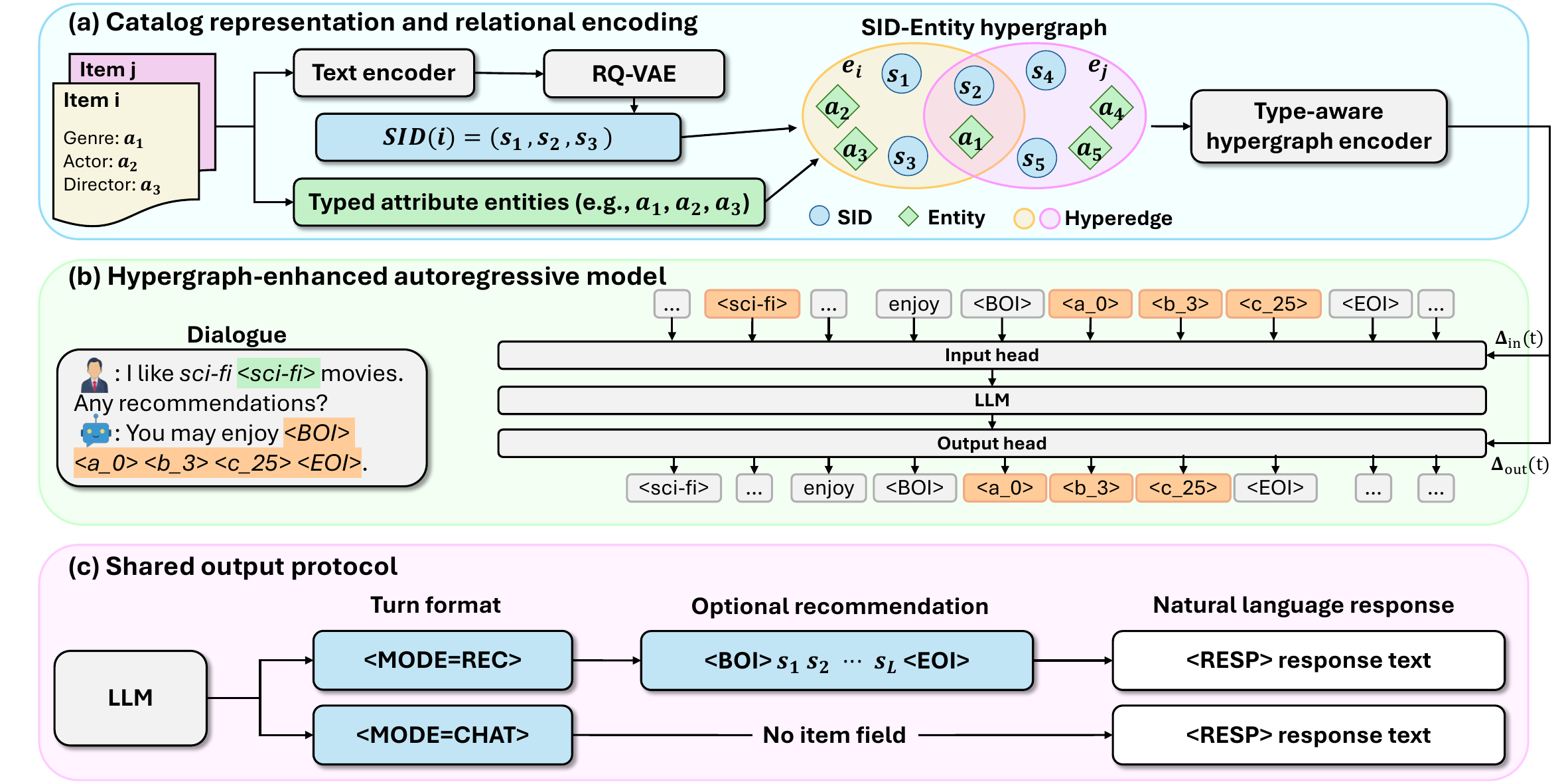}
\caption{Overview of HGCRS. (a) Item metadata is converted into semantic IDs,
and each item forms a hyperedge connecting its SID codes and attribute
entities. (b) A type-aware hypergraph encoder produces relational
representations that augment the input embeddings of entity and SID tokens
as well as the output embeddings of SID tokens. (c) The model generates a
mode token, an optional recommendation SID, and a natural-language response.}
\label{fig:gcrs}
\end{figure*}

HGCRS is a unified generative CRS that grounds compositional SID-based item
generation in structured catalog relations. As illustrated in
\autoref{fig:gcrs}, we first construct item SIDs and then build an SID--entity hypergraph whose item hyperedges connect
SID codes with attribute entities. A jointly trained
type-aware hypergraph encoder maps this structure to relational residuals
that augment entity and SID input representations and SID output
representations. HGCRS additionally uses a shared mode--item--response
generation protocol so that recommendation-bearing and response-only
assistant turns can be modeled within the same autoregressive framework.


\subsection{Task Definition}
\label{sec:task_definition}

Conversational recommendation aims to help users discover relevant items
through natural-language interaction. Let
$\mathcal{D}=(u_1,\ldots,\allowbreak u_T)$ denote a dialogue containing $T$
utterances, where $u_\nu$ is the utterance at dialogue position
$\nu\in\{1,\ldots,T\}$. For an assistant turn at position
$t\in\{1,\ldots,T\}$, let
$C_t=(u_1,\ldots,u_{t-1})$ denote its preceding dialogue context,
$\mathcal{I}$ the item set, $i_t\in\mathcal{I}\cup\{\varnothing\}$ the
target item, and $r_t$ the natural-language response. Here,
$i_t=\varnothing$ indicates that the turn introduces no new recommendation.

The CRS task is to learn a parameterized model $f_{\Theta}$, with trainable
parameter set $\Theta$, that predicts an optional recommendation target and
a natural-language response:
\begin{equation}
(\hat{i}_t,\hat{r}_t)
=
f_{\Theta}(C_t,\mathcal{I}),
\qquad
\hat{i}_t\in\mathcal{I}\cup\{\varnothing\},
\label{eq:crs_task}
\end{equation}
where $\hat{i}_t$ and $\hat{r}_t$ denote the predicted recommendation target
and generated response, respectively. If an original assistant turn contains
multiple recommended items, we decompose it into multiple item-specific
instances that share the same dialogue context and response, with each
instance associated with one target item.

\subsection{Semantic ID Construction via RQ-VAE}
\label{sec: SID construction}
We replace all the textual titles of the items in the dialogues with their SIDs \cite{NEURIPS2023_20dcab0f}. Compared with title-only references or assigning
a unique vocabulary token to every item, SIDs provide explicit item
identifiers while sharing a compact code vocabulary
\cite{he2023large,he2025reindex}.

\textbf{Metadata encoding.}
For each item $i\in\mathcal{I}$, we serialize its available metadata into a
textual description $x_i$ and obtain an item representation
$\mathbf{v}_i=\operatorname{Enc}_{\mathrm{text}}(x_i)$ using a fixed
pretrained text encoder $\operatorname{Enc}_{\mathrm{text}}$. An RQ-VAE
quantizes $\mathbf{v}_i$ into $L$ discrete codes, with one code selected from
each of $L$ codebooks of size $K$.

\textbf{Code-token vocabulary.}
Let $j\in\{1,\ldots,L\}$ index the SID position and
$c\in\{0,\ldots,K-1\}$ index a code within a codebook. We write
$c_{i,j}$ for the code assigned to item $i$ at position $j$, and introduce a
dedicated vocabulary token $s_{j,c}$ for code $c$ at that position. The SID
of item $i$ is therefore
\begin{equation}
\mathrm{SID}(i)=
\left(s_{1,c_{i,1}},\ldots,s_{L,c_{i,L}}\right).
\label{eq:sid_definition}
\end{equation}
For example, when $L=4$, an SID may be written as
$\texttt{<a\_17>}$ $\texttt{<b\_63>}$ $\texttt{<c\_0>}$ $\texttt{<d\_25>}$. The SID vocabulary contains $LK$
position-specific code tokens and can represent up to $K^L$ code
combinations. However, residual quantization can assign the same code sequence to multiple items, and
we resolve such collisions following
\cite{zhai2025multimodalquantitativelanguagegenerative}, as detailed in
\autoref{appendix: collision resolution}. The resulting SIDs are
unique for every item and remain fixed during conversational model training.

\subsection{Relation-Informed Recommendation with an SID--Entity Hypergraph}
\label{sec:sid_entity_hypergraph}

The representations used by conventional knowledge-aware recommenders and
SID-based generators operate at different granularities: the former reason
over atomic item nodes, whereas the latter select an item through multiple
shared code tokens. We therefore construct a hypergraph whose nodes operate
at the same SID-code and entity granularity exposed to the language model.

\textbf{Hypergraph construction.}
Let $\mathcal{G}=(\mathcal{V},\mathcal{H})$ denote the SID--entity
hypergraph, where $\mathcal{V}$ is the node set and $\mathcal{H}$ is the set
of item hyperedges. The nodes comprise all the SID codes and attribute entities. The entity inventory is constructed offline from catalog metadata by
collecting the attribute entities associated with all items. Let $\mathcal{E}_{\mathrm{type}}$ denote the set
of entity types (e.g., genre, actor) included in the catalog schema, and let $v_{j,c}\in
\mathcal{V}$ denote the node corresponding to code $c$ at SID position $j$.
For an entity type $\kappa\in\mathcal{E}_{\mathrm{type}}$, let
$\mathcal{A}_{\kappa}(i)\subseteq\mathcal{V}$ denote the set of entity nodes of type
$\kappa$ associated with item $i$. Each item $i$ contributes one hyperedge
$e_i\in\mathcal{H}$ which contains nodes of its SID tokens and related entities:
\begin{equation}
\begin{aligned}
e_i ={}& \{v_{j,c_{i,j}}:j=1,\ldots,L\}\\
&\cup \bigcup_{\kappa\in\mathcal{E}_{\mathrm{type}}}\mathcal{A}_{\kappa}(i).
\end{aligned}
\label{eq:sid_entity_hyperedge}
\end{equation}
Shared entity and code nodes connect different item hyperedges,
allowing relational information to propagate between items through their
attributes and SIDs.

\textbf{Entity tokens in dialogue text.}
For each entity $a$ retained in the entity inventory, we introduce
a dedicated vocabulary token $\tau_a$. If a matched mention of entity $a$
has surface text form $w$, we preserve the surface text and append its entity
token, yielding $w\,\Vert\,\tau_a$, where $\Vert$ denotes sequence
concatenation (e.g., $\texttt{sci-fi}$ followed by $\texttt{<sci-fi>}$ in \autoref{fig:gcrs}(b)). This provides a canonical token-level reference to the
corresponding hypergraph entity while retaining the original linguistic
expression. The transformation is applied to dialogue contexts and target
responses. Entity inventory construction and mention matching are described in \autoref{app:entity_tokens}.

\textbf{Type-aware hypergraph encoding.}
Standard HGNN \cite{feng2019hypergraph} propagates information through normalized
node--hyperedge--node aggregation, while HyperGCN \cite{yadati2019hypergcn}
approximates each hyperedge with pairwise graph connections. Their standard formulations do not distinguish
SID positions from entity types in their feature transformations. In our
hypergraph, however, an item contains exactly one code from each SID position
but can contain multiple entities of the same type. We therefore use
role-specific transformations and balance the contribution of different
roles during aggregation.

Each node $v\in\mathcal{V}$ is initialized with a trainable representation
$\mathbf{h}_v^{(0)}\in\mathbb{R}^{d_g}$, where $d_g$ is the hypergraph
hidden dimension. Let $L_g$ denote the number of
layers and let $\ell\in\{0,\ldots,L_g-1\}$ index a layer. We define the set
of aggregation roles as
\begin{equation}
\mathcal{R}=
\{\mathrm{SID}_1,\ldots,\mathrm{SID}_L\}
\cup \mathcal{E}_{\mathrm{type}},
\label{eq:hypergraph_roles}
\end{equation}
where the first $L$ roles correspond to SID positions and the remaining
roles correspond to entity types. Let
$\rho:\mathcal{V}\rightarrow\mathcal{R}$ map each node to its role. For a
hyperedge $e\in\mathcal{H}$ and role $r\in\mathcal{R}$, define
\begin{equation}
\mathcal{V}_e^r=
\{v\in e:\rho(v)=r\},
\qquad
\mathcal{R}(e)=
\{r\in\mathcal{R}:|\mathcal{V}_e^r|>0\},
\label{eq:hypergraph_role_sets}
\end{equation}
where $\mathcal{V}_e^r$ is the set of nodes of role $r$ incident to $e$ and
$\mathcal{R}(e)$ is the set of nonempty roles in that hyperedge.

At layer $\ell$, we first aggregate nodes into each hyperedge. Let
$\mathbf{a}_e^{(\ell)}\in\mathbb{R}^{d_g}$ denote the role-balanced
aggregate:
\begin{equation}
\mathbf{a}_e^{(\ell)}=
\frac{1}{|\mathcal{R}(e)|}
\sum_{r\in\mathcal{R}(e)}
\frac{1}{|\mathcal{V}_e^r|}
\sum_{v\in\mathcal{V}_e^r}
W_{r,\mathrm{N}\to\mathrm{E}}^{(\ell)}\mathbf{h}_v^{(\ell)},
\label{eq:sid_hyperedge_aggregation}
\end{equation}
where $W_{r,\mathrm{N}\to\mathrm{E}}^{(\ell)}\in\mathbb{R}^{d_g\times d_g}$ is the
role-specific node-to-hyperedge transformation for role $r$. The hyperedge
representation is then updated as
\begin{equation}
\mathbf{z}_e^{(\ell+1)}=
\operatorname{GELU}\!\left(
W_{\mathrm{edge}}^{(\ell)}
\operatorname{LN}\!\left(\mathbf{a}_e^{(\ell)}\right)
+\mathbf{b}_{\mathrm{edge}}^{(\ell)}
\right),
\label{eq:sid_hyperedge_update}
\end{equation}
where $\mathbf{z}_e^{(\ell+1)}\in\mathbb{R}^{d_g}$ is the updated
representation of hyperedge $e$,
$W_{\mathrm{edge}}^{(\ell)}\in\mathbb{R}^{d_g\times d_g}$ and
$\mathbf{b}_{\mathrm{edge}}^{(\ell)}\in\mathbb{R}^{d_g}$ are trainable
parameters, $\operatorname{LN}$ denotes layer normalization, and
$\operatorname{GELU}$ denotes the GELU activation. The nested averages in
\autoref{eq:sid_hyperedge_aggregation} assign each nonempty role the same
explicit aggregation weight regardless of how many nodes of that role occur
in the hyperedge.

The updated hyperedge representations are then propagated back to their
incident nodes:
\begin{equation}
\mathbf{h}_v^{(\ell+1)}=
\operatorname{LN}\!\left(
\mathbf{h}_v^{(\ell)}
+
W_{\rho(v),\mathrm{E}\to\mathrm{N}}^{(\ell)}
\frac{1}{\deg(v)}
\sum_{\substack{e\in\mathcal{H}\\v\in e}}
\mathbf{z}_e^{(\ell+1)}
\right),
\label{eq:sid_hypernode_update}
\end{equation}
where
$W_{\rho(v),\mathrm{E}\to\mathrm{N}}^{(\ell)}\in\mathbb{R}^{d_g\times d_g}$ is the
role-specific hyperedge-to-node transformation for node $v$, and
$\deg(v)=|\{e\in\mathcal{H}:v\in e\}|$ is the number of hyperedges incident
to $v$. After $L_g$ propagation layers, the final representation of
node $v$ is
\begin{equation}
\mathbf{g}_v=\mathbf{h}_v^{(L_g)}.
\label{eq:final_graph_representation}
\end{equation}

\textbf{Input and output embedding residuals.}
The entity node embeddings as well as the SID node embeddings are then used to augment the corresponding LLM embeddings. Let $\mathcal{T}_{\mathrm{ent}}$ and $\mathcal{T}_{\mathrm{sid}}$
denote the entity and SID tokens associated with the hypergraph. Each such token corresponds to exactly one hypergraph node. Let
\[
\phi:
\mathcal{T}_{\mathrm{ent}}\cup\mathcal{T}_{\mathrm{sid}}
\rightarrow \mathcal{V}
\]
be the token-to-node mapping. Let $d_{\mathrm{LM}}$ denote the LLM embedding dimension. We use separate trainable bridges for the
input and output interfaces. For
$\eta\in\{\mathrm{in},\mathrm{out}\}$, let
$P_\eta\in\mathbb{R}^{d_{\mathrm{LM}}\times d_g}$ and
$\mathbf{b}_\eta\in\mathbb{R}^{d_{\mathrm{LM}}}$ denote the corresponding
projection matrix and bias, $\alpha_\eta\in\mathbb{R}$ a trainable scalar
gate logit, $s_\eta>0$ a fixed scale factor, and $\sigma(\cdot)$ the sigmoid
function. The hypergraph residual for token $t\in\mathcal{T}_{\mathrm{ent}}\cup\mathcal{T}_{\mathrm{sid}}$ at interface $\eta$ is
\begin{equation}
\boldsymbol{\Delta}_{\eta}(t)=
s_{\eta}\sigma(\alpha_{\eta})
\operatorname{LN}\!\left(
P_{\eta}{\mathbf{g}}_{\phi(t)}+\mathbf{b}_{\eta}
\right).
\label{eq:sid_hypergraph_residual}
\end{equation}

Let $E_{\mathrm{in}}(t)\in\mathbb{R}^{d_{\mathrm{LM}}}$ denote the ordinary
LLM input embedding of token $t$. For
$t\in\mathcal{T}_{\mathrm{ent}}\cup\mathcal{T}_{\mathrm{sid}}$, its effective
input embedding is
\begin{equation}
\widetilde{E}_{\mathrm{in}}(t)=
E_{\mathrm{in}}(t)+\boldsymbol{\Delta}_{\mathrm{in}}(t).
\label{eq:sid_hypergraph_input}
\end{equation}
These residuals augment entity- and SID-token occurrences in dialogue
contexts and ground-truth target sequences. All other input tokens use
their ordinary embeddings.

Let $E_{\mathrm{out}}(t)\in\mathbb{R}^{d_{\mathrm{LM}}}$ denote the ordinary
output-head embedding for token $t$. For
$t\in\mathcal{T}_{\mathrm{sid}}$, we define
\begin{equation}
\widetilde{E}_{\mathrm{out}}(t)=
E_{\mathrm{out}}(t)+\boldsymbol{\Delta}_{\mathrm{out}}(t).
\label{eq:sid_hypergraph_output}
\end{equation}

Let $\mathbf{q}\in\mathbb{R}^{d_{\mathrm{LM}}}$ denote the LLM decoding
state at a prediction step. The hypergraph-enhanced logit for SID token $t$ is
then computed as
\begin{equation}
\widetilde{\ell}_{t}
=
\mathbf{q}^{\top}\widetilde{E}_{\mathrm{out}}(t)
\label{eq:sid_hypergraph_logit}
\end{equation}
Thus, hypergraph information augments the existing SID output representation
rather than replacing its ordinary trainable output embedding. Entity-token
and ordinary-text output embeddings receive no hypergraph residual. The fixed
scales $s_{\mathrm{in}}$ and $s_{\mathrm{out}}$ calibrate the hypergraph residuals
to the magnitude of the corresponding native LLM embedding spaces, while
the trainable gates control the contribution of relational information at
the input and output interfaces. The gates, projection modules, and
hypergraph representations are optimized jointly during conversational
fine-tuning. The ordinary embeddings of newly introduced entity and SID
tokens remain trainable alongside these hypergraph residuals.

\subsection{Shared Output Protocol via Structured Generation}
\label{sec:structured_generation}

To model heterogeneous assistant behaviors within the same autoregressive
framework, HGCRS uses a shared sequence format for both
recommendation-bearing and response-only turns.

\textbf{Item representation in dialogue text.}
For each item $i$, we replace its textual title with its SID and delimit the item span using the beginning-of-item token
$\texttt{<BOI>}$ and end-of-item token $\texttt{<EOI>}$:
\begin{equation}
\operatorname{Item}(i)=
\texttt{<BOI>}\,\mathrm{SID}(i)\,\texttt{<EOI>}.
\label{eq:marked_item}
\end{equation}
Together with the entity-token augmentation described above, this
preprocessing transforms the original context $C_t$ and response $r_t$ into
a processed context $\widetilde{C}_t$ and processed response
$\widetilde{r}_t$, respectively.

\textbf{Output structure.}
Let $m_t\in{\mathrm{REC},\mathrm{CHAT}}$ denote the mode associated with
training instance $t$, where $m_t=\mathrm{REC}$ if $i_t\in\mathcal{I}$ and
$m_t=\mathrm{CHAT}$ if $i_t=\varnothing$. The vocabulary contains the
corresponding control tokens $\texttt{<MODE=REC>}$ and
$\texttt{<MODE=CHAT>}$; let $M(m_t)$ denote the control token associated
with mode $m_t$. We use $\texttt{<RESP>}$ to mark the beginning of the
natural-language response segment.

Let $Q_t$ denote the optional recommendation segment and $Y_t$ the complete
supervised assistant sequence for instance $t$. We define
\begin{equation}
\begin{aligned}
Y_t
&=
M(m_t)\,\Vert\,Q_t\,\Vert\,\texttt{<RESP>}\,\Vert\,\widetilde{r}_t,\\
Q_t
&=
\begin{cases}
\operatorname{Item}(i_t), & m_t=\mathrm{REC},\\
\varepsilon, & m_t=\mathrm{CHAT},
\end{cases}
\end{aligned}
\label{eq:structured_sequences}
\end{equation}
where $\Vert$ denotes concatenation and $\varepsilon$ denotes the
empty token sequence. A recommendation-bearing instance therefore generates
the SID of its target item before the response, whereas a response-only
instance proceeds directly from the mode token to $\texttt{<RESP>}$.

Let $P_{\Theta}$ denote the autoregressive distribution parameterized by the
trainable parameter set $\Theta$. The structured sequence induces the conditional factorization
\begin{equation}
\begin{aligned}
P_{\Theta}(Y_t\mid\widetilde{C}_t)
={}&
P_{\Theta}(m_t\mid\widetilde{C}_t)\,
P_{\Theta}(Q_t\mid m_t,\widetilde{C}_t)\\
&\times
P_{\Theta}(\widetilde{r}_t\mid Q_t,m_t,\widetilde{C}_t),
\end{aligned}
\label{eq:factorization}
\end{equation}
This factorization makes the generation stages
explicit: the mode determines whether the current instance introduces a new
recommendation, the optional recommendation segment specifies the target
item when required, and the response conditions on the preceding structured
fields. All stages are realized through autoregressive decoding in the same
language model.

\subsection{Training and Inference}
\label{sec:training_inference}

\textbf{Joint next-token training.}
For each training instance, let $|Y_t|$ denote the number of tokens in the
supervised assistant sequence $Y_t$. Let
$p\in\{1,\ldots,|Y_t|\}$ index a target position, let $y_{t,p}$ be the
token at that position, and let
$y_{t,<p}=(y_{t,1},\ldots,y_{t,p-1})$ denote its preceding target prefix.
We optimize the next-token prediction objective
\begin{equation}
\mathcal{L}_{\mathrm{NTP}}
=
-\sum_t\sum_{p=1}^{|Y_t|}
\log
P_{\Theta}\!\left(
y_{t,p}\mid
\widetilde{C}_t,y_{t,<p}
\right),
\label{eq:ntp_loss}
\end{equation}
where $\mathcal{L}_{\mathrm{NTP}}$ denotes the next-token prediction
loss over training instances. The loss covers the mode token, recommendation
SID when present, structural delimiters, and response tokens. The trainable
parameter set $\Theta$ includes the hypergraph node representations, hypergraph
message-passing parameters, input and output bridges and gates, newly
introduced entity, SID, and control-token embeddings, and the
parameter-efficient LLM adaptation parameters.

\textbf{Structured decoding.}
Given a processed dialogue context $\widetilde{C}_t$, the model first
generates or receives a mode token and then completes the sequence specified
by that mode. The mode can either be generated
automatically from the dialogue context or supplied as a prefix to request a
particular response behavior. During recommendation decoding, a trie built from the
complete set of catalog SIDs restricts each generated SID prefix to one that
can be completed into a valid catalog identifier. The trie constrains only
the recommendation segment, while natural-language response generation
remains unconstrained.

\section{Experiments}
\label{sec: experiments}

\begin{table*}[t]
\centering
\caption{Recommendation performance on ReDial and INSPIRED (\%).
The best overall result is \textbf{bolded}, and the second-best result is \underline{underlined}. $^*$ denotes statistical
significance ($p<0.05$) over other methods using a paired
$t$-test. Improv. reports the relative improvement of HGCRS over the
strongest baseline.}
\vspace{-1em}
\label{table:rec}
\small
\begin{tabular}{ll*{10}{r}}
\toprule
Dataset & Model & R@1 & R@5 & R@10 & R@20 & N@5 & N@10 & N@20 & M@5 & M@10 & M@20 \\
\midrule
\multirow{11}{*}{ReDial}
& KGSF   & 3.52 & 11.87 & 17.91 & 26.18 & 7.63 & 9.59 & 11.67 & 6.25 & 7.06 & 7.62 \\
& MESE   & 4.10 & 13.50 & 21.01 & 29.45 & 8.82 & 11.24 & 13.38 & 7.29 & 8.28 & 8.87 \\
& DCRS   & 5.12 & 15.19 & 21.93 & 29.65 & 10.20 & 12.39 & 14.33 & 8.56 & 9.47 & 10.00 \\
& STEP   & 5.57 & 15.36 & 22.36 & 30.85 & 10.59 & 12.85 & 15.00 & 9.02 & 9.95 & 10.54 \\
& UniCRS & 4.66 & 14.85 & 21.46 & 29.62 & 9.83 & 11.95 & 14.01 & 8.18 & 9.04 & 9.61 \\
& MSCRS  & 5.65 & 16.04 & 23.01 & 30.48 & 10.84 & 13.10 & 14.99 & 9.15 & 10.08 & 10.60 \\
& TalkPlay
& \underline{6.58} & \underline{16.92} & \underline{23.73}
& \underline{31.08} & \underline{11.74} & \underline{13.55}
& \underline{15.48} & \underline{10.06} & \underline{10.96}
& \underline{11.49} \\
& TIGER  & 2.69 & 11.16 & 17.83 & 25.41 & 6.92 & 9.06 & 10.97 & 5.53 & 6.41 & 6.93 \\
& LC-Rec & 3.18 & 10.74 & 17.28 & 24.54 & 7.00 & 9.08 & 10.90 & 5.78 & 6.61 & 7.11 \\
\cmidrule(lr){2-12}
& \textbf{HGCRS}
& $\textbf{7.17}^{*}$ & $\textbf{18.30}^{*}$ & $\textbf{25.22}^{*}$
& $\textbf{33.06}^{*}$ & $\textbf{12.85}^{*}$ & $\textbf{15.07}^{*}$
& $\textbf{17.05}^{*}$ & $\textbf{11.06}^{*}$ & $\textbf{11.97}^{*}$
& $\textbf{12.51}^{*}$ \\
& Improv.
& \textcolor{ForestGreen}{+8.97\%}
& \textcolor{ForestGreen}{+8.16\%}
& \textcolor{ForestGreen}{+6.28\%}
& \textcolor{ForestGreen}{+6.37\%}
& \textcolor{ForestGreen}{+9.45\%}
& \textcolor{ForestGreen}{+11.22\%}
& \textcolor{ForestGreen}{+10.14\%}
& \textcolor{ForestGreen}{+9.94\%}
& \textcolor{ForestGreen}{+9.22\%}
& \textcolor{ForestGreen}{+8.88\%} \\

\midrule
\multirow{11}{*}{INSPIRED}
& KGSF   & 5.18 & 13.27 & 15.21 & 17.80 & 9.34 & 9.98 & 10.64 & 8.05 & 8.32 & 8.50 \\
& MESE   & 3.88 & 11.00 & 14.24 & 20.71 & 7.88 & 8.92 & 10.57 & 6.82 & 7.25 & 7.70 \\
& DCRS   & 4.30 & 10.55 & 16.80 & 21.88 & 7.11 & 9.09 & 10.36 & 6.02 & 6.82 & 7.15 \\
& STEP   & 7.03 & 17.58 & 22.27 & 27.34 & 12.67 & 14.20 & 15.51 & 11.04 & 11.68 & 12.05 \\
& UniCRS & 9.77 & 19.53 & 23.83 & 30.86 & 14.69 & 16.03 & 17.77 & 13.11 & 13.63 & 14.09 \\
& MSCRS  & 10.16 & 20.70 & 27.73 & 33.20 & 15.46 & 17.77 & 19.14 & 13.74 & 14.72 & 15.09 \\
& TalkPlay
& \underline{12.56} & \underline{23.37} & \underline{29.45}
& \underline{35.61} & \underline{19.25} & \underline{20.87}
& \underline{22.08} & \underline{17.11} & \underline{18.02}
& \underline{19.06} \\
& TIGER  & 11.28 & 16.92 & 22.93 & 29.70 & 14.27 & 16.15 & 17.91 & 13.39 & 14.13 & 14.64 \\
& LC-Rec & 9.02 & 17.29 & 22.18 & 25.94 & 13.21 & 14.76 & 15.71 & 11.87 & 12.50 & 12.76 \\
\cmidrule(lr){2-12}
& \textbf{HGCRS}
& $\textbf{14.89}^{*}$ & $\textbf{27.83}^{*}$ & $\textbf{33.01}^{*}$
& $\textbf{38.84}^{*}$ & $\textbf{21.74}^{*}$ & $\textbf{23.39}^{*}$
& $\textbf{24.87}^{*}$ & $\textbf{19.72}^{*}$ & $\textbf{20.39}^{*}$
& $\textbf{20.79}^{*}$ \\
& Improv.
& \textcolor{ForestGreen}{+18.55\%}
& \textcolor{ForestGreen}{+19.08\%}
& \textcolor{ForestGreen}{+12.09\%}
& \textcolor{ForestGreen}{+9.07\%}
& \textcolor{ForestGreen}{+12.94\%}
& \textcolor{ForestGreen}{+12.07\%}
& \textcolor{ForestGreen}{+12.64\%}
& \textcolor{ForestGreen}{+15.25\%}
& \textcolor{ForestGreen}{+13.15\%}
& \textcolor{ForestGreen}{+9.08\%} \\
\bottomrule
\end{tabular}
\vspace{-1em}
\end{table*}

We evaluate HGCRS through the following research questions:
\begin{itemize}
    \item RQ1: How does HGCRS compare with existing methods in recommendation accuracy and response generation?
    \item RQ2: How do relational grounding and the key design choices affect recommendation and response generation?
    \item RQ3: How sensitive is HGCRS to the choice of LLM backbone and metadata encoder?
\end{itemize}

\textbf{Datasets.}
We use \textbf{ReDial} \cite{li2018towards} and \textbf{INSPIRED}
\cite{hayati2020inspired}, two benchmarks containing multi-turn movie
recommendation dialogues. Dataset statistics are provided in
\autoref{appendix:datasets}.

\textbf{Baselines.}
We compare against two groups of methods. The CRS baselines are
\textbf{KGSF} \cite{zhou2020improving}, \textbf{MESE}
\cite{yang2021improving}, \textbf{DCRS} \cite{dao2024broadening},
\textbf{STEP} \cite{yang2025step}, \textbf{UniCRS}
\cite{wang2022towards}, \textbf{MSCRS} \cite{wei2025mscrs}, and \textbf{TalkPlay} \cite{doh2025talkplay}. The generative recommendation baselines are
\textbf{TIGER} \cite{NEURIPS2023_20dcab0f} and \textbf{LC-Rec}
\cite{10597986}. For a controlled comparison, HGCRS and all baselines use
\texttt{\hspace{0pt}Qwen2.5-\allowbreak 7B-\allowbreak Instruct}
\cite{qwen2.5} as the backbone. We compare all methods on
recommendation accuracy and report response-generation metrics for methods
that generate natural-language responses.

\textbf{Evaluation metrics.}
For recommendation, we report \textbf{Recall@$k$}, \textbf{NDCG@$k$}, and
\textbf{MRR@$k$}, abbreviated as R@$k$, N@$k$, and M@$k$, respectively.
Recommendation metrics are computed on turns whose reference responses
contain recommendations. Following TIGER \cite{NEURIPS2023_20dcab0f}, we
generate a ranked item list through constrained beam search over the SID
space starting from the first \texttt{<BOI>} token, and the top-50 valid items are ranked by their beam scores for
computing the ranking metrics. For comparability with CRS baselines that always return a ranked list, the
main recommendation evaluation conditions HGCRS on the reference turn mode.
Only the binary mode token is supplied; no target item, SID, or response
information is revealed. We do not evaluate autonomous mode selection by
exact agreement with the reference mode, because the recorded response
represents only one plausible conversational trajectory rather than a
uniquely correct action. For example, we observe a test dialogue in which the
user says ``Thanks. Goodbye'', while the recorded assistant response
recommends another movie; both another recommendation and an appropriate
farewell can be conversationally reasonable. We therefore evaluate
autonomous mode selection through the human \textbf{mode appropriateness} metric described below, which
assesses whether the automatically generated response behavior is appropriate for the given dialogue context.

For response generation, we report \textbf{perplexity (PPL)},
\textbf{SacreBLEU} \cite{post2018call}, and \textbf{Distinct-$n$} (D-$n$).
Before computing the metrics, structural control tokens and dedicated
entity tokens are removed while the original entity surface forms are
retained. Item spans represented by
\texttt{<BOI>}\,$\mathrm{SID}$\,\texttt{<EOI>} are replaced with a common
\texttt{<movie>} placeholder in both predictions and references. All baselines share the same setting. Automatic metrics are averaged over five independent runs,
and statistical significance is assessed using a paired $t$-test at
$p<0.05$. Detailed definitions are provided in
\autoref{appendix:metrics}. 

We further conduct human evaluation along three dimensions:
\textbf{fluency}, \textbf{informativeness}, and \textbf{mode
appropriateness}. The annotation protocol and results are reported in
\autoref{appendix: human eval}.

\textbf{Implementation.}
By default, we instantiate HGCRS with \texttt{\hspace{0pt}Qwen 2.5-\allowbreak 7B-\allowbreak Instruct} \cite{qwen2.5} and encode item metadata using \texttt{Sentence-\allowbreak T5} \cite{ni2022sentence} for ReDial and \texttt{BGE} \cite{bge_embedding} for INSPIRED. We train the model using parameter-efficient fine-tuning. Detailed model
configurations and optimization settings are provided in \autoref{appendix:implementation}.

\begin{table*}[t]
\caption{Response generation performance on ReDial and INSPIRED.
Lower PPL is better, while higher BLEU and Distinct-$n$ are better.
The best and second-best results are \textbf{bolded} and
\underline{underlined}, respectively. $^*$ denotes statistical significance
($p<0.05$) over other methods using a paired $t$-test.}
\vspace{-1em}
\label{table:redial dialog}
\small
\setlength{\tabcolsep}{3.5pt}
\centering
\begin{tabular}{l*{12}{r}}
\toprule
 & \multicolumn{6}{c}{ReDial} & \multicolumn{6}{c}{INSPIRED} \\
\cmidrule(lr){2-7} \cmidrule(lr){8-13}
 & PPL & BLEU & D-1 & D-2 & D-3 & D-4
 & PPL & BLEU & D-1 & D-2 & D-3 & D-4 \\
\midrule
KGSF
& 237.68 & \textbf{26.65} & 0.0020 & 0.0504 & 0.1852 & 0.3513
& 464.84 & \textbf{11.82} & 0.0019 & 0.0153 & 0.0395 & 0.0624 \\

MESE
& 8.48 & 4.53 & 0.0218 & 0.0927 & 0.1814 & 0.2633
& 15.44 & 5.23 & 0.0730 & 0.2200 & 0.3385 & 0.4068 \\

DCRS
& 12.14 & 9.52 & \textbf{0.0383} & \textbf{0.1443}
& \textbf{0.2638} & \underline{0.3806}
& 21.13 & 5.27 & 0.0826 & 0.2378 & 0.3899 & 0.5195 \\

STEP
& 85.67 & 3.70 & 0.0023 & 0.0112 & 0.0781 & 0.1544
& 18.14 & 5.58 & 0.1080 & 0.3168 & 0.4988 & 0.6280 \\

UniCRS
& 34.00 & 11.14 & 0.0169 & 0.0603 & 0.1186 & 0.1880
& 32.43 & 4.40 & 0.0659 & 0.2009 & 0.3386 & 0.4577 \\

MSCRS
& 10.09 & 8.98 & \underline{0.0353} & 0.1278 & 0.2374 & 0.3483
& 15.55 & 5.71 & 0.0968 & 0.2821 & 0.4513 & 0.5771 \\

TalkPlay
& \underline{4.46} & 11.96 & 0.0327 & 0.1298 & 0.2417 & 0.3584
& \underline{6.46} & 5.76 & \underline{0.1295}
& \underline{0.3691} & \underline{0.5399} & \underline{0.6401} \\

\cmidrule(lr){1-13}

\textbf{HGCRS}
& $\textbf{4.40}$ & \underline{12.19} & 0.0329
& \underline{0.1349} & \underline{0.2615} & $\textbf{0.3871}$
& $\textbf{5.63}^{*}$ & \underline{5.99}
& $\textbf{0.1315}$ & $\textbf{0.3711}$
& $\textbf{0.5519}^{*}$ & $\textbf{0.6712}^{*}$ \\
\bottomrule
\end{tabular}
\vspace{-1em}
\end{table*}

\begin{figure*}[t]
    \centering
    \begin{subfigure}[t]{0.49\textwidth}
        \centering
        \includegraphics[width=\linewidth]{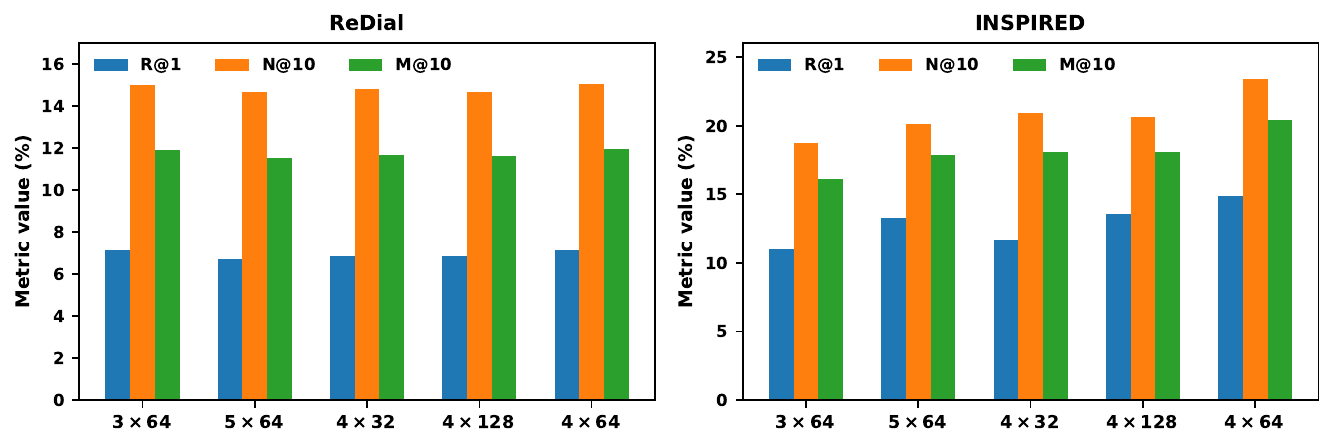}
        \caption{SID configuration ablations.}
        \label{fig:ablation_sid}
    \end{subfigure}
    \hfill
    \begin{subfigure}[t]{0.49\textwidth}
        \centering
        \includegraphics[width=\linewidth]{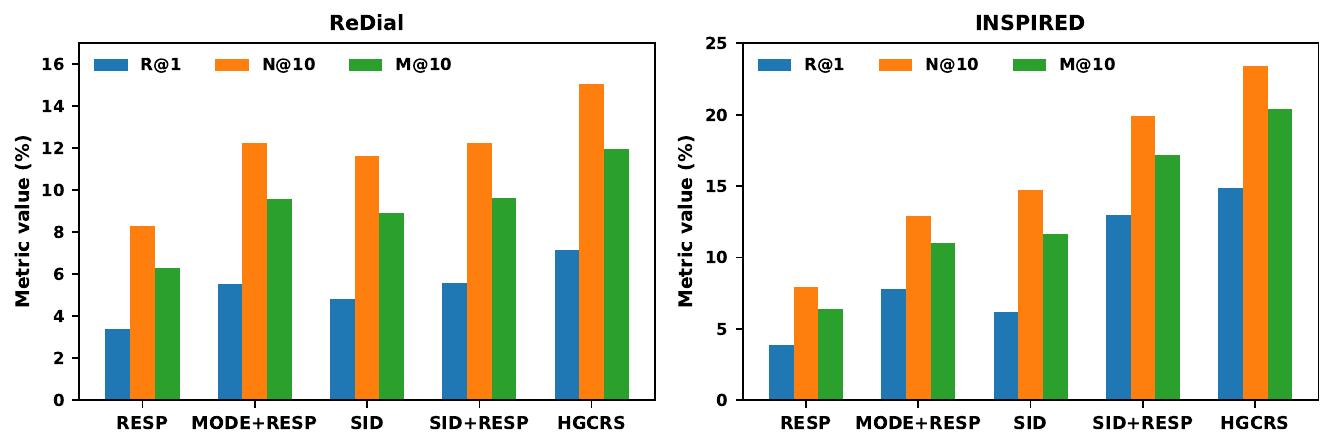}
        \caption{Structured output ablations.}
        \label{fig:ablation_output}
    \end{subfigure}
    \vspace{-1em}
    \caption{Ablations of SID configurations and structured output supervision
    on ReDial and INSPIRED.}
    \label{fig:ablation_design}
    \vspace{-1em}
\end{figure*}

\subsection{RQ1: Overall Performance}

\subsubsection{Recommendation Performance}
\autoref{table:rec} reports the recommendation results on ReDial and
INSPIRED. HGCRS achieves the strongest performance across all reported
ranking metrics on both datasets and significantly outperforms the strongest baseline, demonstrating its effectiveness for conversational item
recommendation. Among the baselines, TalkPlay provides the strongest overall recommendation
performance, outperforming the other CRS baselines as well as the SID-based
generative recommenders. Nevertheless,
HGCRS consistently surpasses TalkPlay across all ranking metrics. Since TalkPlay
already jointly models recommended items and natural-language responses,
the remaining gains therefore cannot be attributed solely to adopting a
unified generative formulation. HGCRS further grounds generative item
selection in structured catalog relations aligned with the representations
used by the generator. RQ2 isolates the contributions of this relational
grounding and the associated design choices.

\subsubsection{Response Generation Performance}
\autoref{table:redial dialog} shows that HGCRS maintains strong response
generation performance. We additionally conduct human evaluation, with the
results and annotation protocol reported in
\autoref{appendix: human eval}. HGCRS achieves the lowest PPL and the
second-best BLEU on both datasets. It also shows strong lexical diversity,
achieving the best D-4 on ReDial and the best D-1 through D-4 on INSPIRED.
KGSF obtains the highest BLEU despite substantially lower diversity,
highlighting the limitation of reference-overlap metrics in capturing
response diversity. Since entity tokens are excluded and item mentions are replaced
by a shared placeholder during evaluation, the diversity gains cannot be
attributed to the enlarged vocabulary. Together with the human evaluation,
these results show that the recommendation gains of HGCRS do not come at the
cost of response quality.

\begin{figure}[t]
    \centering
    \includegraphics[width=\columnwidth]{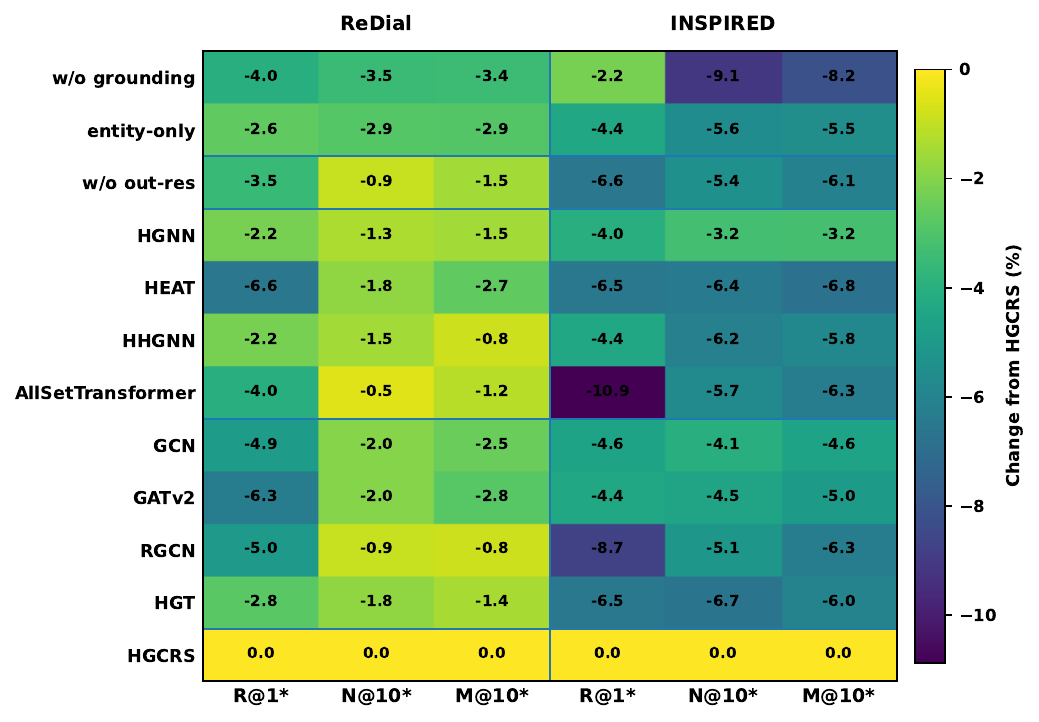}
    \vspace{-2em}
    \caption{Ablations of HGCRS. Each cell shows the relative
    change (\%) from the complete HGCRS configuration. $^*$ marks metrics for
    which HGCRS significantly outperforms all variants ($p<0.05$).}
    \label{fig:ablation_core}
    \vspace{-1em}
\end{figure}

\begin{table*}[t]
\centering
\caption{Backbone and metadata-encoder comparisons on ReDial and INSPIRED.
Backbone experiments use Sentence-T5 and BGE as metadata encoders for ReDial and INSPIRED, respectively, while
metadata-encoder experiments use Qwen2.5 as the backbone. The best and second-best
results in each comparison are \textbf{bolded} and \underline{underlined}, respectively.}
\label{table:llms}
\vspace{-1em}
\small
\setlength{\tabcolsep}{2.2pt}
\begin{tabular}{l*{14}{r}}
\toprule
& \multicolumn{7}{c}{ReDial}
& \multicolumn{7}{c}{INSPIRED} \\
\cmidrule(lr){2-8}\cmidrule(lr){9-15}
& R@1 & R@10 & N@10 & M@10 & PPL & BLEU & D-2
& R@1 & R@10 & N@10 & M@10 & PPL & BLEU & D-2 \\
\midrule
\multicolumn{15}{l}{\textit{(a) Backbone comparison}} \\
Llama
& \underline{7.13} & 24.89 & 14.49 & 11.36 & 4.52 & 11.84 & 0.1119
& \underline{12.62} & 31.39 & 21.31 & 18.18 & 5.80 & \underline{6.15} & 0.3306 \\
Qwen3
& 6.47 & 24.30 & 14.50 & 11.48 & \textbf{4.37} & \underline{12.04} & 0.1053
& \underline{12.62} & \underline{32.69} & \underline{22.24} & \underline{18.92}
& 5.93 & 5.75 & 0.2959 \\
Mistral
& \underline{7.13} & \underline{25.20} & \textbf{15.09} & \underline{11.96}
& 4.56 & 11.98 & \textbf{0.1460}
& 11.97 & 32.04 & 21.33 & 17.99 & \textbf{5.09} & \textbf{6.29} & \underline{0.3419} \\
Ministral-3
& 7.11 & 24.26 & 14.68 & 11.74 & 4.58 & 11.52 & 0.1230
& \underline{12.62} & 31.39 & 21.71 & 18.63 & 5.66 & 5.70 & 0.2648 \\
Qwen2.5
& \textbf{7.17} & \textbf{25.22} & \underline{15.07} & \textbf{11.97}
& \underline{4.40} & \textbf{12.19} & \underline{0.1349}
& \textbf{14.89} & \textbf{33.01} & \textbf{23.39} & \textbf{20.39}
& \underline{5.63} & 5.99 & \textbf{0.3711} \\
\midrule
\multicolumn{15}{l}{\textit{(b) Metadata-encoder comparison}} \\
Sentence-T5
& \underline{7.17} & \textbf{25.22} & \textbf{15.07} & \textbf{11.97}
& 4.40 & \textbf{12.19} & 0.1349
& 12.30 & \underline{30.10} & \underline{20.87} & 17.95
& 5.67 & 5.87 & 0.3679 \\
E5
& \underline{7.17} & 24.32 & 14.61 & 11.65 & 4.44 & 12.04 & 0.1318
& 12.30 & 29.13 & 20.37 & 17.61
& \underline{5.65} & \underline{5.93} & 0.3596 \\
Llama
& 6.88 & 24.77 & 14.68 & 11.59 & \textbf{4.37} & \underline{12.13} & 0.1352
& \underline{12.95} & 28.80 & 20.84 & \underline{18.32}
& 5.66 & 5.86 & \textbf{0.3771} \\
Mxbai
& \underline{7.17} & 24.24 & 14.67 & 11.74
& \underline{4.38} & 11.95 & \underline{0.1360}
& 12.62 & \underline{30.10} & 20.50 & 17.54
& 5.76 & 5.64 & 0.3442 \\
BGE
& \textbf{7.19} & \underline{25.17} & \underline{15.05} & \underline{11.93}
& 4.40 & 12.00 & \textbf{0.1376}
& \textbf{14.89} & \textbf{33.01} & \textbf{23.39} & \textbf{20.39}
& \textbf{5.63} & \textbf{5.99} & \underline{0.3711} \\
\bottomrule
\end{tabular}
\vspace{-1.5em}
\end{table*}

\subsection{RQ2: Ablation and Design Analysis}
\autoref{fig:ablation_core} examines relational grounding, residual
integration, and relational encoders, while \autoref{fig:ablation_sid} and
\autoref{fig:ablation_output} analyze SID configurations and structured
output supervision. For clarity, we only show recommendation metrics in these figures. Complete response-generation results are provided in
\autoref{appendix:ablation}. We additionally examine the effects of different item representations and SID construction strategies in \autoref{appendix:item_representation}.

\textbf{Relational grounding and residual integration.}
We first examine how relational information is introduced into the generator in \autoref{fig:ablation_core}.
The \emph{w/o grounding} variant removes both dedicated entity tokens and the
SID--entity hypergraph, reducing HGCRS to the SID-based structured generator,
whereas \emph{entity-only} retains trainable entity tokens but removes all hypergraph residuals. Entity tokens alone provide only modest and inconsistent gains over the basic SID-based generator, while the complete HGCRS
consistently improves recommendation performance over both variants. This
indicates that the improvement cannot be attributed merely to exposing
explicit entity identifiers to the LLM; propagating catalog relations between
entity and SID-code nodes provides additional recommendation signal. The \emph{w/o out-res} variant removes the hypergraph residual applied to SID output
representations. Its consistent degradation shows
that input-side relational grounding alone does not fully exploit the encoded
catalog structure. Directly augmenting SID output representations helps
translate relational information into the scores used for generative item
selection.

\textbf{Graph and hypergraph encoders.}
In \autoref{fig:ablation_core}, we further replace the proposed type-aware encoder with existing hypergraph
encoders (HGNN \cite{feng2019hypergraph}, HEAT \cite{georgiev2022heat}, HHGNN \cite{li2022heterogeneous}, and AllSetTransformer \cite{chien2021you}) and conventional graph
encoders (GCN \cite{kipf2016semi}, GATv2 \cite{brody2021attentive}, RGCN \cite{schlichtkrull2018modeling}, and HGT \cite{hu2020heterogeneous}), while keeping the remaining
components unchanged. For conventional graph encoders, each item hyperedge is converted into
pairwise connections among its incident SID and entity nodes using clique
expansion. The two encoder families are broadly comparable on ReDial, while the strongest hypergraph
alternatives show a clearer advantage on INSPIRED. Nevertheless, generic
hypergraph modeling does not consistently match HGCRS, suggesting that
explicitly distinguishing SID positions and entity types provides additional
benefit.

\textbf{Semantic ID configurations.}
\autoref{fig:ablation_sid} examines the performance of different SID configurations. The default $4\times64$ configuration
provides the strongest overall ranking performance. ReDial is relatively
robust to SID configuration, whereas INSPIRED is more sensitive, particularly
to shorter SIDs and smaller codebooks. We therefore use $4\times64$ as the
default configuration.

\textbf{Structured generation components.}
\autoref{fig:ablation_output} ablates the three fields of the proposed
MODE+SID+RESP protocol by modifying both the training target and the
corresponding inference procedure. RESP retains only the response segment,
discarding both the mode token and the explicit target-item segment.
MODE+RESP retains the mode token and response segment but removes the
target-item segment. SID retains only the target-item segment and discards
both the mode token and textual response, whereas SID+RESP jointly supervises
the target-item and response segments while removing the mode token. Recommendation inference is adapted to the output structure of each variant.
For RESP, generation starts from \texttt{<RESP>} and proceeds without an
explicit target-item prefix. After the complete response is generated, we
parse the output for a \texttt{<BOI>}SID\texttt{<EOI>} item span. If such
a span is present, we perform a decoding pass from the corresponding \texttt{<BOI>} token using constrained beam search to get the top-$k$ metrics. If no such span is produced
for a recommendation-bearing test instance, the instance is counted as a
miss. MODE+RESP follows the same two-stage procedure, except that generation
is initialized with the reference mode token, consistent with the
reference-mode evaluation used in the main experiments. In contrast, SID
starts directly from \texttt{<BOI>} and applies constrained beam search
to generate the target SID, while SID+RESP uses the same constrained SID
decoding and subsequently continues to generate the response segment.

These variants isolate complementary roles of the structured output fields
under their corresponding inference protocols. The complete HGCRS achieves
the strongest overall recommendation performance, supporting the explicit
separation of recommendation decision, item selection, and conversational
response within a single autoregressive interface.

\subsection{RQ3: Choice of Backbone and Metadata Encoder}
\autoref{table:llms} compares different LLM backbones and metadata encoders.
For the backbone comparison, we retain the default metadata encoder for each
dataset; for the encoder comparison, we fix the backbone to Qwen2.5. The
default configuration therefore participates in both comparisons.

\textbf{Backbone LLMs.}
We compare \texttt{Llama-3.1-8B} \cite{grattafiori2024llama},
\texttt{Qwen3-8B} \cite{qwen3technicalreport}, \texttt{Mistral-7B}
\cite{jiang2023mistral7b}, \texttt{Ministral-3-8B}
\cite{liu2026ministral}, and \texttt{Qwen2.5-7B} \cite{qwen2.5}.
No single backbone dominates every metric,
although Qwen2.5 provides the strongest overall recommendation performance,
particularly on INSPIRED. Other backbones remain competitive on individual
ranking or generation metrics. These results show that the structured
generation and relational-integration mechanisms of HGCRS can be applied across
different LLM families rather than depending on a particular backbone.

\textbf{Metadata encoders for SID construction.}
With Qwen2.5 fixed, we compare \texttt{Sentence-T5} \cite{ni2022sentence},
\texttt{BGE} \cite{bge_embedding}, \texttt{E5} \cite{wang2022text},
\texttt{Llama} \cite{touvron2023llama}, and \texttt{Mxbai}
\cite{emb2024mxbai} as metadata encoders. Encoder choice has a relatively
small effect on ReDial but a more noticeable impact on INSPIRED, suggesting
that the preferred semantic representation is dataset dependent. Overall,
HGCRS remains reasonably robust across the evaluated encoders. Since the
metadata encoder determines the SID assignments, changing it simultaneously
changes the discrete item representations and the SID-code incidences in the
hypergraph; the observed differences therefore reflect both effects.



\section{Conclusion}
We presented HGCRS, which aligns structured catalog relations with
SID-based generative CRS through an SID--entity hypergraph and
type-aware relational propagation. A structured mode--item--response protocol
further unifies recommendation decisions, item selection, and response
generation. Experiments on ReDial and INSPIRED demonstrate consistent
recommendation gains while maintaining competitive response quality.
\bibliographystyle{ACM-Reference-Format}
\bibliography{sample-base}

\appendix

\section{Collision Resolution}
\label{appendix: collision resolution}
Suppose a group of $N$ items is initially assigned the same SID. For each
colliding item $i$, quantization level $j\in\{1,\ldots,L\}$, and codeword
$k\in\{0,\ldots,K-1\}$, we compute the squared Euclidean distance between
the residual representation $\mathbf{r}_i^{(j)}$ and codeword
$\mathbf{e}_k^{(j)}$:
\begin{equation}
d_{i,k}^{(j)}
=
\left\|
\mathbf{r}_i^{(j)}
-
\mathbf{e}_k^{(j)}
\right\|_2^2.
\end{equation}
These distances form a tensor
\begin{equation}
\mathbf{D}\in\mathbb{R}^{N\times L\times K}.
\end{equation}
For each item and quantization level, candidate codewords are ordered by
increasing distance. We first rank the colliding items according to their nearest-codeword
distance at the final quantization level, giving priority to items with more
confident assignments. Starting from the final level, each item is assigned
its nearest codeword that yields an available code sequence. If the nearest
choice would preserve an existing collision, the next closest candidate is
considered. When modifying the final-level code is insufficient to resolve
all collisions, the procedure backtracks to the preceding quantization
level, selects an alternative code according to the same distance ordering,
and subsequently reassigns the later levels. This process continues
recursively toward earlier levels when necessary until all colliding items
are assigned distinct code sequences.

\section{Entity Inventory and Mention Matching}
\label{app:entity_tokens}

We construct the entity inventory from the selected attribute fields in the
item metadata. Each attribute value is treated as a catalog entity
and assigned a dedicated entity token. To identify entity mentions in
dialogues, we match the attribute text against dialogue contexts and target
responses using both exact and fuzzy matching. Before matching, entity names and dialogue text are normalized by case
folding, accent removal, and whitespace and punctuation normalization. Exact
matching is applied first and takes precedence over fuzzy matches. Fuzzy
matching is used to account for minor surface-form variations and
typographical differences. Specifically, we compare normalized strings using
optimal-string-alignment edit distance and accept candidates with distance at
most one, corresponding to a single insertion, deletion, substitution, or
adjacent transposition. When multiple candidates overlap, exact matches are
preferred; among fuzzy matches, candidates with smaller edit distance and
longer matched spans are prioritized. Ambiguous ties between different
entities are discarded. The resulting entity inventory and matching rules are fixed after catalog
preprocessing and are applied consistently to the training and evaluation
sets.

\section{Datasets}
\label{appendix:datasets}
ReDial contains 10,006 training dialogues and 1,342 test dialogues, covering 6,924 unique items. INSPIRED contains 900 training dialogues and 99 test dialogues, covering 1,782 items. For each dialogue, we construct training samples by treating each assistant utterance as the target response and the preceding dialogue context as input. Movie metadata (title, year, keywords, genres, actors, directors, plot) is collected from IMDb\footnote{\url{https://www.imdb.com/}} and serialized into textual descriptions for SID construction. The hypergraphs are constructed with the SID codes and the entities from genres, actors, and directors. The hypergraph contains 38,058 nodes and 124,947 node-hyperedge incidences for ReDial; 12,967 nodes and 32,054 node-hyperedge incidences for INSPIRED.

\begin{table}[t]
\setlength{\tabcolsep}{2pt}
\centering
\caption{Hyperparameter settings used in the main experiments.}
\vspace{-1em}
\label{tab:hyper-parameter}
\small
\begin{tabular}{lll}
\toprule
         & ReDial & INSPIRED \\
\midrule
\multicolumn{3}{l}{\textit{RQ-VAE}} \\
\midrule
learning rate & 1e-3 & 1e-3 \\
weight decay & 1e-4 & 1e-4 \\
batch size & 1,024 & 1,024 \\
encoder layers & 7 & 7 \\
encoder output size & 32 & 32 \\
quantization levels & 4 & 4 \\
codebook size & 64 & 64 \\
\midrule
\multicolumn{3}{l}{\textit{Hypergraph encoder}} \\
\midrule
encoder layers & 1 & 1 \\
hidden dimension & 128 & 128 \\
\midrule
\multicolumn{3}{l}{\textit{Conversational fine-tuning}} \\
\midrule
learning rate & 2e-4 & 1e-4 \\
batch size & 72 & 72 \\
warmup steps & 150 & 50 \\
training steps & 1,800 & 240 \\
maximum input length & 768 & 768 \\
weight decay & 0.0 & 0.0 \\
LoRA rank & 16 & 16 \\
LoRA alpha & 32 & 32 \\
LoRA dropout & 0.05 & 0.05 \\
recommendation beam width & 50 & 50 \\
\bottomrule
\end{tabular}
\vspace{-1.5em}
\end{table}

\section{Dialogue evaluation metrics}
\label{appendix:metrics}
We evaluate the generated responses from both the language quality and diversity perspectives using Perplexity (PPL), SacreBLEU, and Distinct-$n$.

\subsection{Perplexity}
Perplexity measures how well a probabilistic model predicts a sequence. Given a sequence of tokens $y = (y_1, y_2, \dots, y_T)$, the sequence-level perplexity is defined as:
\begin{equation}
\mathrm{PPL}(y) = \exp\left( - \frac{1}{T} \sum_{t=1}^{T} \log p(y_t \mid y_{<t}) \right).
\end{equation}

For a corpus $\mathcal{D}$ consisting of multiple sequences, we compute the corpus-level perplexity by aggregating over all tokens:
\begin{equation}
\mathrm{PPL}(\mathcal{D}) = \exp\left(
- \frac{\sum_{i=1}^{N} \sum_{t=1}^{T_i} \log p(y_{i,t} \mid y_{i,<t})}
{\sum_{i=1}^{N} T_i}
\right),
\end{equation}
where $N$ is the number of sequences, and $T_i$ is the length of the $i$-th sequence. This formulation corresponds to the exponentiated average negative log-likelihood per token across the entire corpus, ensuring that longer sequences contribute proportionally more to the final score.

\subsection{BLEU}
We evaluate response-reference overlap using corpus-level SacreBLEU
\cite{post2018call}. We use the default English configuration provided by
the SacreBLEU implementation: case-sensitive evaluation with the 13a
tokenizer, BLEU-4, and exponential smoothing. The score is
\begin{equation}
\mathrm{BLEU}
=
\mathrm{BP}\cdot
\exp\left(
\frac{1}{4}\sum_{n=1}^{4}\log p_n
\right),
\end{equation}
where $p_n$ is the corpus-level clipped $n$-gram precision and
\begin{equation}
\mathrm{BP} =
\begin{cases}
1, & c \ge r,\\
\exp\left(1-\frac{r}{c}\right), & c < r,
\end{cases}
\end{equation}
with $c$ and $r$ denoting the cumulative tokenized candidate and reference
lengths, respectively. We report SacreBLEU on its standard 0--100 scale.

\subsection{Distinct-$n$}
Distinct-$n$ measures the diversity of generated text by calculating the ratio of unique $n$-grams over the total number of generated $n$-grams:
\begin{equation}
\mathrm{Distinct}\text{-}n = \frac{|\mathcal{G}_n^{\text{unique}}|}{|\mathcal{G}_n|},
\end{equation}
where $\mathcal{G}_n$ is the set of all $n$-grams in the generated corpus, and $\mathcal{G}_n^{\text{unique}}$ is the set of distinct $n$-grams. We report Distinct-$n$ for $n \in \{1,2,3,4\}$.

\begin{table}[t]
\centering
\caption{Human evaluation on ReDial and INSPIRED. All metrics are rated from 0 to 2.
Higher is better. The best and second-best results are \textbf{bolded} and
\underline{underlined}, respectively.}
\vspace{-1em}
\label{tab:human_eval}
\small
\setlength{\tabcolsep}{3pt}
\begin{tabular}{lcccccc}
\toprule
& \multicolumn{3}{c}{ReDial}
& \multicolumn{3}{c}{INSPIRED} \\
\cmidrule(lr){2-4}\cmidrule(lr){5-7}
Model
& Fluency & Inform. & Mode App.
& Fluency & Inform. & Mode App. \\
\midrule
KGSF
& 1.39 & 1.10 & 1.21
& 1.37 & 1.08 & 1.27 \\
MESE
& 1.62 & 1.22 & 1.32
& 1.60 & 1.20 & 1.32 \\
DCRS
& 1.70 & 1.25 & 1.30
& \underline{1.73} & 1.22 & 1.33 \\
STEP
& 1.67 & 1.39 & 1.35
& 1.50 & \underline{1.33} & 1.39 \\
UniCRS
& 1.70 & 1.29 & 1.28
& 1.67 & 1.13 & 1.20 \\
MSCRS
& 1.72 & 1.33 & 1.35
& 1.63 & 1.18 & 1.29 \\
TalkPlay
& \underline{1.78} & \underline{1.43} & \underline{1.45}
& 1.69 & \underline{1.33} & \underline{1.43} \\
\midrule
\textbf{HGCRS}
& \textbf{1.86} & \textbf{1.53} & \textbf{1.52}
& \textbf{1.84} & \textbf{1.47} & \textbf{1.50} \\
\bottomrule
\end{tabular}
\vspace{-1em}
\end{table}

\begin{table}[t]
\centering
\caption{Response-generation performance of the ablation variants on ReDial
and INSPIRED. Lower PPL is better, while higher BLEU and D-2 are better.
The best and second-best results are \textbf{bolded} and
\underline{underlined}, respectively. \textemdash{} denotes metrics not
applicable to SID-only generation.}
\vspace{-1em}
\label{tab:ablation_response}
\small
\setlength{\tabcolsep}{5.0pt}
\begin{tabular}{ll*{6}{r}}
\toprule
& \multicolumn{3}{c}{ReDial}
& \multicolumn{3}{c}{INSPIRED} \\
\cmidrule(lr){2-4}\cmidrule(lr){5-7}
& PPL & BLEU & D-2
& PPL & BLEU & D-2 \\
\midrule
 w/o grounding
& 4.43 & 11.58 & 0.1216
& 6.36 & 5.75 & 0.3701 \\
entity-only
& 4.39 & 11.99 & 0.1370
& \underline{5.63} & 5.79 & 0.3774 \\
 w/o out-res
& 4.39 & 12.06 & 0.1378
& 5.65 & 5.59 & \underline{0.3776} \\
\midrule
 HGNN
& 4.39 & 12.03 & 0.1344
& 5.65 & 5.54 & \textbf{0.3778} \\
HEAT
& 4.39 & \underline{12.25} & 0.1343
& 5.64 & 5.75 & 0.3753 \\
HHGNN
& 4.38 & 12.04 & \textbf{0.1437}
& 6.07 & 5.48 & 0.3680 \\
AllSetTransformer
& 4.38 & 12.13 & 0.1402
& 5.64 & 5.54 & 0.3701 \\
GCN
& 4.39 & 12.10 & 0.1387
& 5.66 & 5.83 & 0.3742 \\
GATv2
& 4.39 & 12.09 & 0.1407
& 5.66 & 5.84 & 0.3706 \\
RGCN
& 4.40 & 11.97 & 0.1374
& 5.64 & \underline{5.99} & 0.3737 \\
HGT
& 4.39 & 12.06 & 0.1352
& 5.66 & 5.64 & 0.3613 \\
\midrule
 $3\times64$
& 4.74 & 12.20 & 0.1375
& 5.74 & 5.87 & 0.3676 \\
$5\times64$
& \textbf{4.08} & 12.15 & 0.1302
& 5.76 & 5.91 & 0.3504 \\
$4\times32$
& \underline{4.35} & 12.06 & 0.1348
& \textbf{5.58} & \textbf{6.16} & 0.3660 \\
$4\times128$
& 4.44 & 12.07 & 0.1332
& 5.71 & 5.83 & 0.3554 \\
\midrule
 RESP
& 6.59 & 10.11 & \underline{0.1418}
& 7.30 & 4.63 & 0.3594 \\
MODE+RESP
& 5.58 & \textbf{12.39} & 0.1396
& 6.52 & 4.94 & 0.3573 \\
SID+RESP
& 4.67 & 12.09 & 0.1367
& 6.03 & 5.84 & 0.3692 \\
SID
& \textemdash & \textemdash & \textemdash
& \textemdash & \textemdash & \textemdash \\
\midrule
 \textbf{HGCRS}
& 4.40 & 12.19 & 0.1349
& \underline{5.63} & \underline{5.99} & 0.3711 \\
\bottomrule
\end{tabular}
\vspace{-1.5em}
\end{table}

\begin{table*}[t]
\centering
\caption{Item-representation and SID-construction results on ReDial and INSPIRED. The best and second-best available
results are \textbf{bolded} and \underline{underlined}, respectively. $^*$ denotes statistical significance ($p<0.05$) over other methods using a paired $t$-test.}
\vspace{-1em}
\label{tab:item_representation}
\small
\setlength{\tabcolsep}{4pt}
\begin{tabular}{lcccc}
\toprule
\multirow{2}{*}{Item representation}
& \multicolumn{2}{c}{Recommendation}
& \multicolumn{2}{c}{Response generation} \\
\cmidrule(lr){2-3}\cmidrule(lr){4-5}
& INSPIRED R@1/5/10/20
& ReDial R@1/5/10/20
& INSPIRED PPL/BLEU/D-2
& ReDial PPL/BLEU/D-2 \\
\midrule
Unique token
& 7.44/15.86/20.39/23.95
& 4.94/13.97/20.36/28.12
& 8.18/5.57/0.3687
& 6.88/\underline{11.87}/0.1314 \\
Textual title
& 8.41/20.06/26.86/33.33
& 5.06/12.69/17.35/24.63
& \underline{5.97}/\textbf{6.41}/0.3414
& \underline{4.42}/11.48/\textbf{0.1381} \\
SID (collaborative)
& \underline{13.57}/24.21/30.09/33.45
& 6.56/16.98/23.02/31.24
& 6.22/5.56/0.3701
& 4.94/11.53/0.1325 \\
SID (metadata + collaborative)
& 13.16/\underline{25.49}/\underline{31.22}/\underline{34.27}
& \underline{6.75}/\underline{17.44}/\underline{23.61}/\underline{31.74}
& 6.22/\underline{5.99}/\textbf{0.3840}
& 4.58/11.03/0.1218 \\
HGCRS
& $\textbf{14.89}^{*}$/$\textbf{27.83}^{*}$/$\textbf{33.01}^{*}$/$\textbf{38.84}^{*}$
& $\textbf{7.17}^{*}$/$\textbf{18.30}^{*}$/$\textbf{25.22}^{*}$/$\textbf{33.06}^{*}$
& \textbf{5.63}/\underline{5.99}/\underline{0.3711}
& \textbf{4.40}/\textbf{12.19}/\underline{0.1349} \\
\bottomrule
\end{tabular}
\vspace{-1.5em}
\end{table*}

\section{Implementation Details}
\label{appendix:implementation}
The main hyperparameters are summarized in
\autoref{tab:hyper-parameter}. We construct SIDs from item metadata representations using RQ-VAE. \texttt{Sentence-T5} \cite{ni2022sentence} is used as the metadata
encoder for ReDial and \texttt{BGE} \cite{bge_embedding} for INSPIRED. The RQ-VAE is
trained independently before conversational fine-tuning, and the resulting
item-to-SID assignments remain fixed during subsequent training. Details of
SID collision handling are provided in \autoref{appendix: collision resolution}. The default language-model backbone is
\texttt{Qwen2.5-7B-Instruct} \cite{qwen2.5}. We apply QLoRA to the linear
layers. The original vocabulary embedding rows are kept fixed, while newly
introduced SID, entity, and control-token embeddings are trainable. Hypergraph node embeddings, the hypergraph encoder,
the input/output projection and gating modules, newly introduced token
embeddings, and the QLoRA parameters are optimized jointly during
conversational fine-tuning. For the hypergraph residual defined in
\autoref{eq:sid_hypergraph_residual}, the input and output scale factors
$s_{\mathrm{in}}$ and $s_{\mathrm{out}}$ are calibrated separately from the
corresponding pretrained embedding tables. Specifically, each scale is set
to the estimated median norm of the native-token embeddings divided by
$\sqrt{d_{\mathrm{LM}}}$, matching the magnitude of the layer-normalized
hypergraph residual to that of the native embedding space. The input and output
gate logits are initialized as
$\alpha_{\mathrm{in}}=\alpha_{\mathrm{out}}=-4$, corresponding to an
initial gate value of $\sigma(-4)\approx0.018$, and are subsequently learned
jointly with other trainable parameters. Experiments are conducted on a single NVIDIA A100 80GB PCIe GPU.

\section{Human Evaluation}
\label{appendix: human eval}
We conduct human evaluation to assess response quality beyond
automatic metrics. We randomly sample 100 instances from the test
set of each dataset and ask three annotators to independently evaluate the
generated responses. All responses are generated without reference-mode conditioning. Model identities are anonymized and the response order is randomized. For each instance, annotators are shown the dialogue
context together with the outputs from all compared models and assign scores
along three dimensions: \textbf{fluency}, \textbf{informativeness}, and
\textbf{mode appropriateness}. All are rated on a 0--2 scale. Fluency measures whether a response is
grammatically correct, coherent, and natural in the given conversational
context, while informativeness measures whether it provides useful and
relevant information rather than a generic or uninformative reply. Mode appropriateness evaluates whether the model makes an appropriate
interaction-level decision for the current dialogue context: either providing
a recommendation-bearing response or continuing with a chat-only response. This criterion evaluates the contextual appropriateness of the
model's decision rather than exact agreement with the reference turn mode.
The final score for each dimension is averaged across the three annotators
and all sampled instances.

As shown in \autoref{tab:human_eval}, HGCRS achieves the strongest scores on
all three dimensions on both datasets. In particular, its higher mode
appropriateness indicates that the model not only generates fluent and
informative responses, but also more reliably decides whether recommendation
is suitable for the current conversational context. Together with the
automatic results, the human evaluation shows that the unified framework
maintains response quality while making appropriate recommendation decisions.

\section{Response-generation Results of Ablations}
\label{appendix:ablation}
\autoref{tab:ablation_response} complements the recommendation-focused
ablations with response-generation metrics. Relational grounding, residual,
and encoder variants produce comparatively small and non-monotonic changes
in PPL, BLEU, and D-2, suggesting that these components primarily affect
item selection. SID and output variants show larger variation, and the
configuration favored by an individual generation metric does not always
maximize recommendation performance. The default HGCRS configuration
therefore prioritizes robust ranking performance while retaining competitive
response quality.

\section{Analysis of Item Representations and SID Construction Strategies}
\label{appendix:item_representation}

We further investigate how item representation and the information used for
SID construction affect HGCRS. As shown in
\autoref{tab:item_representation}, we consider two types of alternatives.
\emph{Unique token} assigns each item an independent trainable vocabulary
token, while \emph{textual title} directly represents items using their
titles; neither variant uses graph- or hypergraph-based relational
augmentation. We additionally construct alternative SIDs while keeping the
remaining HGCRS architecture unchanged. For \emph{SID (collaborative)}, user
interaction sequences are extracted from the dialogues and encoded with
SASRec \cite{kang2018self} to obtain collaborative item embeddings, which are then quantized by
RQ-VAE in place of metadata embeddings. \emph{SID (Metadata +
collaborative)} combines metadata and collaborative representations following
the SID construction strategy of LETTER \cite{wang2024learnable}.

HGCRS with metadata-based SIDs achieves the best Recall at every cutoff on
both datasets. Collaborative and metadata--collaborative SIDs substantially
outperform the simpler item representations but remain below HGCRS,
indicating that metadata-based SID construction is the most effective choice
for recommendation in our setting. Response-generation results are less
consistent across representations: HGCRS obtains the lowest PPL on both
datasets and the highest BLEU on ReDial, while other representations lead on
individual BLEU or diversity metrics. This suggests that item representation
has a more systematic effect on recommendation accuracy than on response
generation quality.

\end{document}